\documentclass[10pt]{article}
\usepackage{fancyhdr}
\usepackage{extramarks}
\usepackage{amsmath}
\usepackage{amsthm}
\usepackage{amsfonts}
\usepackage{siunitx}
\usepackage{tikz}
\usepackage[plain]{algorithm}
\usepackage{algpseudocode}
\usepackage{multirow}
\usepackage{booktabs}
\usepackage{graphicx}
\usepackage{subfigure}
\usepackage[colorlinks,linkcolor=black,anchorcolor=black,citecolor=black,urlcolor=blue]{hyperref}
\usepackage{amsmath,bm}
\usepackage{booktabs}
\usepackage{mathtools}
\usepackage{amssymb}
\usepackage{caption}
\usepackage{capt-of}
\usepackage{mciteplus}
\usepackage{cite}
\usepackage{mathrsfs}
\usepackage[title,titletoc,toc]{appendix}
\usepackage{xr}
\usepackage{parskip}
\usepackage{soul}
\usepackage{textcomp}
\usepackage[colaction]{multicol}
\usepackage[switch]{lineno}
\usepackage{lipsum}
\usepackage{etoolbox}
\usepackage{longtable}
\usepackage{array}
\usepackage{tablefootnote}
\usepackage{ragged2e}
\usepackage{filecontents}
\usepackage{titling}
\newcolumntype{C}[1]{>{\centering\arraybackslash}p{#1}}
\usetikzlibrary{automata,positioning}
\usetikzlibrary{automata,positioning}

\newcommand{\beginsupplement}{%
    \setcounter{table}{0}
    \renewcommand{\thetable}{S\arabic{table}}%
    \setcounter{figure}{0}
    \renewcommand{\thefigure}{S\arabic{figure}}%
    \setcounter{section}{0}
    \renewcommand{\thesection}{S\arabic{section}}
}

\begin{document}

\title{Omicron BA.2 (B.1.1.529.2): high potential to becoming the next dominating variant
} 
 
\author{ Jiahui Chen$^1$ and Guo-Wei Wei$^{1,3,4}$\thanks{
 		Corresponding author.		Email: weig@msu.edu} \\% Author name
 $^1$ Department of Mathematics, \\
 Michigan State University, MI 48824, USA.\\
 $^2$ Department of Electrical and Computer Engineering,\\
 Michigan State University, MI 48824, USA. \\
 $^3$ Department of Biochemistry and Molecular Biology,\\
 Michigan State University, MI 48824, USA.
}
\date{\today} % Date for the report

\maketitle

\begin{abstract}
The Omicron variant  of severe acute respiratory syndrome coronavirus 2 (SARS-CoV-2) has rapidly replaced the Delta variant as a dominating SARS-CoV-2 variant because of natural selection, which favors the variant with higher infectivity and stronger vaccine breakthrough ability. Omicron has three lineages or subvariants, BA.1 (B.1.1.529.1), BA.2 (B.1.1.529.2), and BA.3 (B.1.1.529.3). Among them, BA.1 is the currently prevailing subvariant. BA.2 shares 32 mutations with BA.1 but has 28 distinct ones. BA.3 shares most of its mutations with BA.1 and BA.2 except for one. BA.2 is found to be able to alarmingly reinfect patients originally infected by Omicron BA.1. An important question is whether BA.2 or BA.3 will become a new dominating ``variant of concern''. Currently, no experimental data has been reported about BA.2 and BA.3. We construct a novel algebraic topology-based deep learning model trained with tens of thousands of mutational and deep mutational data to systematically evaluate BA.2's and BA.3's infectivity, vaccine breakthrough capability, and antibody resistance. Our comparative analysis of all main variants namely, Alpha, Beta, Gamma, Delta, Lambda, Mu, BA.1, BA.2, and BA.3, unveils that BA.2 is about 1.5 and 4.2 times as contagious as BA.1 and Delta, respectively. It is also 30\% and 17-fold more capable than BA.1 and Delta, respectively, to escape current vaccines.  Therefore, we project that Omicron BA.2 is on its path to becoming the next dominating variant. We forecast that like Omicron BA.1, BA.2   will also seriously compromise most existing mAbs, except for sotrovimab developed by   GlaxoSmithKline. 

\end{abstract}
Keywords: COVID-19, SARS-CoV-2, Omicron,  infectivity, antibody-resistance, vaccine breakthrough,  
%\pagenumbering{roman}
%\begin{verbatim}
%\end{verbatim}
%
% {\setcounter{tocdepth}{4} \tableofcontents}
%%
\newpage
 %\clearpage
 %\pagebreak

\setcounter{page}{1}
\renewcommand{\thepage}{{\arabic{page}}}

% \begin{multicols}{2}
% \multicollinenumbers
% \linenumbers

%
 
\section{Introduction}
On November 26, 2021, the World Health Organization (WHO) declared the Omicron variant (B.1.1.529) of severe acute respiratory syndrome coronavirus 2 (SARS-CoV-2)  initially discovered in South Africa a variant of concern (VOC).  Within a few days (i.e., December 1, 2021), an artificial intelligence (AI) model predicted the Omicron variant to be about 2.8 times as infectious as the Delta variant, have a near 90\% likelihood to escape current vaccines, and severely compromise the efficacy of monoclonal antibodies (mAbs) developed by Eli Lilly, Regeneron, AstraZeneca, and many others, except for GlaxoSmithKline’s sotrovimab \cite{chen2022omicron}. The subsequent experiments confirm Omicron's high infectivity \cite{shuai2022attenuated,hong2022molecular}, high vaccine breakthrough rate \cite{cele2021omicron,zhang2022significant}, and severe antibody escape rate \cite{liu2021striking,lu2021neutralization,hoffmann2021omicron}. The U.S. Food and Drug Administration (FDA)  halted the use of mAbs from Eli Lilly and  Regeneron in January 2022. Due to its combined effects of high infectivity and high vaccine breakthrough rate, the Omicron variant is far more transmissible than the Delta variant and has rapidly become the dominating variant in the world. 

Omicron  has three lineages, BA.1 (B.1.1.529.1), BA.2 (B.1.1.529.2), and BA.3 (B.1.1.529.3), which were first detected in November 2021 in South Africa \cite{desingu2022emergence}. 
Among them, BA.1 lineage is the preponderance that has ousted Delta. Compared to the reference genome reported in Wuhan, Omicron BA.1 has a total of 60 mutations on non-structure protein (NSP3), NSP4, NSP5, NSP6, NSP12, NSP14, S protein, envelope protein, membrane protein, and nucleocapsid protein.
 % 50 nonsynonymous mutations, 8 synonymous mutations, and 2 non-coding mutations. 
Among them, 32 mutations are on the spike (S) protein, the main antigenic target of antibodies generated by either infection or vaccination.  
Fifteen of these mutations affect the receptor-binding domain (RBD), whose binding with host angiotensin-converting enzyme 2 (ACE2) facilitates the viral 
cell entry during the initial infection \cite{walls2020structure}. 
% primed by transmembrane serine protease 2 (TMPRSS2) \cite{matsuyama2020enhanced}.
BA.2 shares 32 mutations with BA.1 but has 28 distinct ones. On the RBD, BA.2 has four unique mutations and 12 shared with BA.1. In contrast, the Delta variant has only two RBD mutations. BA.3 shares most of its mutations with BA.1 and BA.2, except for one on NSP6 (A88V). It also has 15 RBD mutations, but none is distinct from BA.1 and BA.2. 
Nationwide Danish data in late December 2021 and early January 2022 indicate that Omicron BA.2 is inherently substantially more transmissible than BA.1 and capable of vaccine breakthrough \cite{lyngse2022transmission}. Israel reported a handful of cases of patients who were infected with original Omicron BA.1  strain and have  reinfected with BA.2 in a short period \cite{timesofisrael}. Although BA.2 did not cause worse illness than the original Omicron BA.1 strain, its reinfection is very alarming. It means the antibodies generated from the early Omicron BA.1 were evaded by the BA.2 strain. It is imperative to know whether BA.2 will become the next dominating strain to reinfect the world population.  

Currently, there are no experimental results about the infectivity, vaccine breakthrough, and antibody resistance of BA.2 and BA.3  \cite{world2021enhancing}. In this work, we present a comprehensive analysis of Omicron BA.2 and BA.3's potential of becoming the next prevailing SARS-CoV-2 variant.    Our study focuses on the S protein RBD, which is essential for virus cell entry. 
Studies show that binding free energy (BFE) between the S RBD and the ACE2 is proportional to the viral infectivity \cite{li2005bats,hoffmann2020sars,walls2020structure}. In July 2020, it was discovered that SARS-CoV-2 evolution is governed by infectivity-based natural selection \cite{chen2020mutations}, which was conformed beyond doubt in April 2021 \cite{wang2021vaccine}. The RBD is not only crucial for viral infectivity but also essential for vaccines and antibody protections. An antibody that can disrupt the RBD-ACE2 binding would directly neutralize the virus \cite{wang2020human,yu2020receptor,li2021impact}.  
We integrate tens of thousands of mutational and deep mutational data, biophysics, and algebraic topology to construct an AI model. 
We systematically investigate the binding free energy (BFE) changes of an RBD-ACE2 complex structure and a library of 185 structures of RBD-antibody complexes induced by the RBD mutations of Alpha, Beta, Gamma, Delta, Lambda, Mu,   BA.1,   BA.2, and   BA.3 to reveal their infectivity, vaccine-escape potential, and antibody resistance. Using our comparative analysis, we unveil  that the Omicron BA.2 variant is about 1.5 times as infectious as BA.1 and about 4.2 times as contagious as the Delta variant. It also has a 30\% higher potential than BA.1 to escape existing vaccines.  Therefore, we project the Omicron BA.2 is on its path to becoming the next dominating variant. 

\begin{figure}[ht!]
	\centering
	\includegraphics[width = 0.94\textwidth]{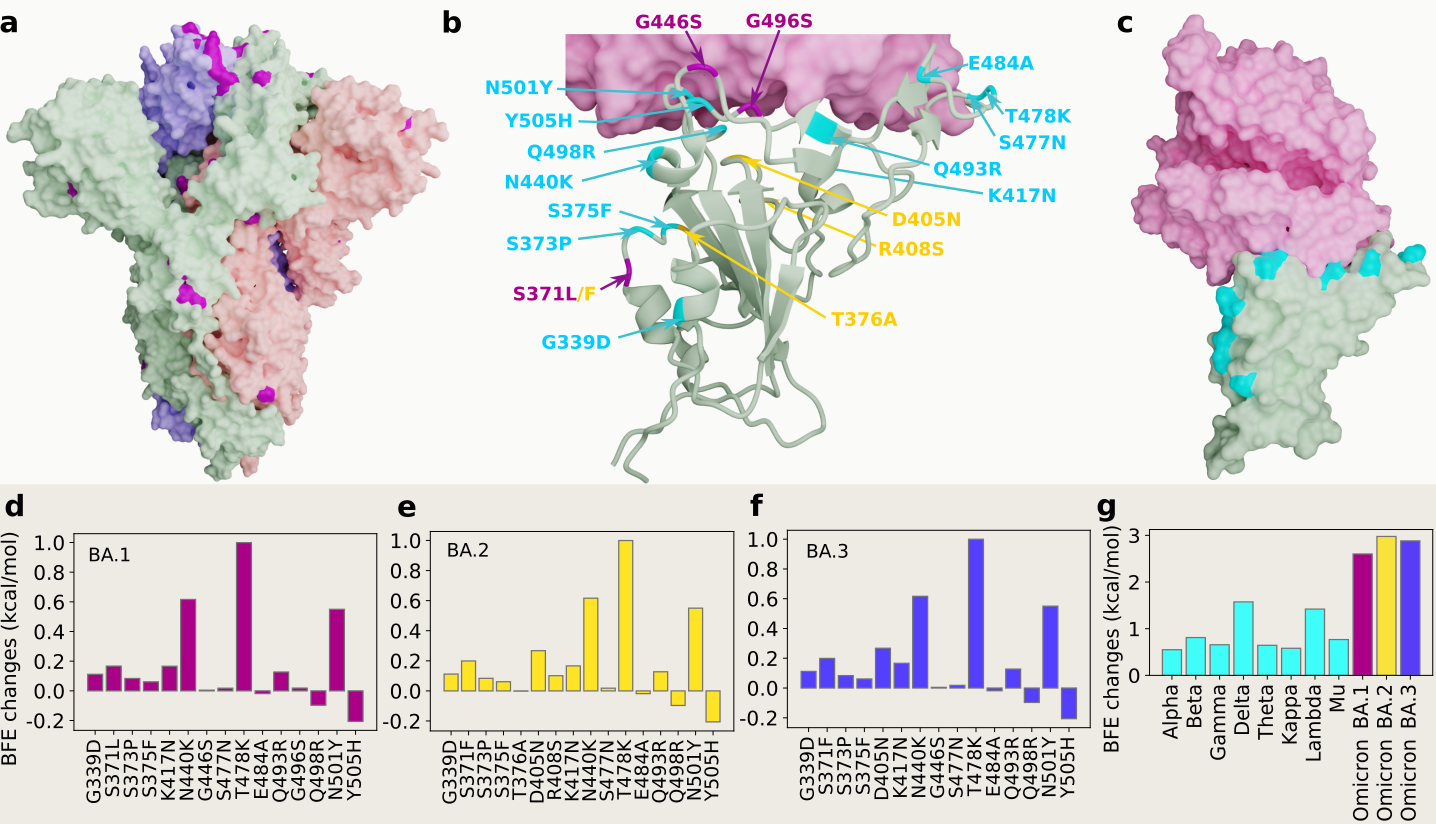}
	\caption{3D structures of Omicron strains, their ACE2 complexes and their mutation-induced BFE changes. {\bf a} Spike protein (PDB: 7WK2 \cite{hong2022molecular}) with Omicron mutations being marked yellow. {\bf b}  BA.1 and BA.2 RBD mutations at the RBD-ACE interface   (PDB: 7T9L \cite{zhu2021cryo}). The shared 12 mutations are   labeled in cyan, BA.1 mutations are marked with magenta, and distinct BA.2 mutations are plotted in yellow. {\bf b} The structure of the RBD-ACE2 complex with mutations on cyan spots. 
{\bf e, f} and {\bf g} BFE changes induced by mutations of Omicron BA.1,  BA.2,  BA.3, respectively.  
{\bf h} a comparison of predicted mutation-induced BFE changes for few SARS-CoV-2 variants.
	}
	\label{fig:combine1}
\end{figure}
\section{Results}
\subsection{Infectivity}
Figure \ref{fig:combine1} {\bf a} shows the three-dimensional (3D) structure of Omicron BA.1 \cite{hong2022molecular}. At the RBD, Omicron BA.1, BA.2 and BA.3 share 12 RBD mutations, i.e., G339D, S373P, S375F, K417N, N440K, S477N, T478K, E484A, Q493R, Q498R, N501Y, and Y505H as shown in Figure \ref{fig:combine1} {\bf b}. However, BA.1 has distinct RBD mutations S371L, G446S, and G496S, BA.2 has  S371F, T376A, D405N, and R408S, and BA.3 has S371F, D405N, and  G446S. Figures \ref{fig:combine1} {\bf d, e} and {\bf f} present the BFE changes of the RBD-ACE2 complex induced by the RBD mutations of Omicron AB.1, BA.2 and BA.3, respectively. The larger the BFE change is, the higher infectivity will be. 
Since natural selection favors those mutations that strengthen the viral infectivity \cite{chen2020mutations}, the most contagious variant will become dominant in a population under the same competing condition.  The accumulated BFE changes are summarized in  Figure  \ref{fig:combine1} {\bf g}. A comparison is given to other main SARS-CoV-2 variants Alpha, Beta, Gamma, Delta, Theta, Kappa, Lambda, and Mu.  The Delta variant had the highest BFE change among the earlier variants and was the most infectious variant before the occurrence of the Omicron variant, which explains its dominance in 2021. Omicron BA.1, BA.2, and BA.3 have BFE changes of 2.60, 2.98, and 2.88  kcal/mol, respectively, which are much higher than those of other major SRAS-CoV-2 variants. Among them, Omicron BA.2  is the most infectious variant and is about 20 and 4.2 times as infectious as the original SARS-CoV-2 and the Delta variant, respectively.  Our model predicts that BA.2 is about 1.5 as contagious BA.2, which is the same as reported in an initial study \cite{timesofisrael}.   Another report confirms that Omicron BA.2  is more contagious than BA.1 \cite{lyngse2022transmission}.  Therefore, Omicron BA.2 may eventually replace the original Omicron strain BA.1 in the world.

\subsection{Vaccine breakthrough}
Omicron BA.1 is well-known for its ability to escape current vaccines \cite{zhang2022significant,liu2021striking}. 
Its 15 mutations at the RBD enable it to not only strengthen its infectivity by a stronger binding to human ACE2 but also create mismatches for most direct neutralization antibodies generated from vaccination or prior infection. Although BA.1, BA.2, and BA.3 share 12 RBD mutations, BA.1 has 3 additional RBD mutations, BA.2 has 4 additional RBD mutations, and BA.3 has one mutation the same as that of BA.1's additional ones and two mutations the same as those of BA.2's additional ones. Therefore, it is important to understand their vaccine-escape potentials.  Currently, no experimental result has been reported about the vaccine-breakthrough capability of BA.2 and BA.3. 

Experimental analysis of the variant vaccine-escape capability over the world's populations is subject to many uncertainties. Different vaccines may stimulate different immune responses and antibodies for the same person. Different individuals may have different immune responses and antibodies from the same vaccine due to their different races, gender, age, and underlying medical conditions. Uncontrollable experimental conditions and different experimental methods may also contribute to uncertainties. Consequently, it is impossible to accurately characterize a variant's vaccine-escape capability (or rate) over the world's populations. 

In our work, we take an integrated approach to understanding the intrinsic vaccine-escape capability of SARS-CoV-2 variants. We collect a library of 185 known antibody and S protein complexes and analyze the mutational impact on the binding of these complexes \cite{chen2021prediction,chen2022omicron}. The results in terms of mutation-induced BFE changes serve as the statistical ensemble analysis of Omicron subvariants' vaccine-breakthrough potentials.  This molecular-level analysis becomes very useful when it is systematically applied to a series of variants. 

\begin{figure}[ht!]
	\centering
	\includegraphics[width = 0.7\textwidth]{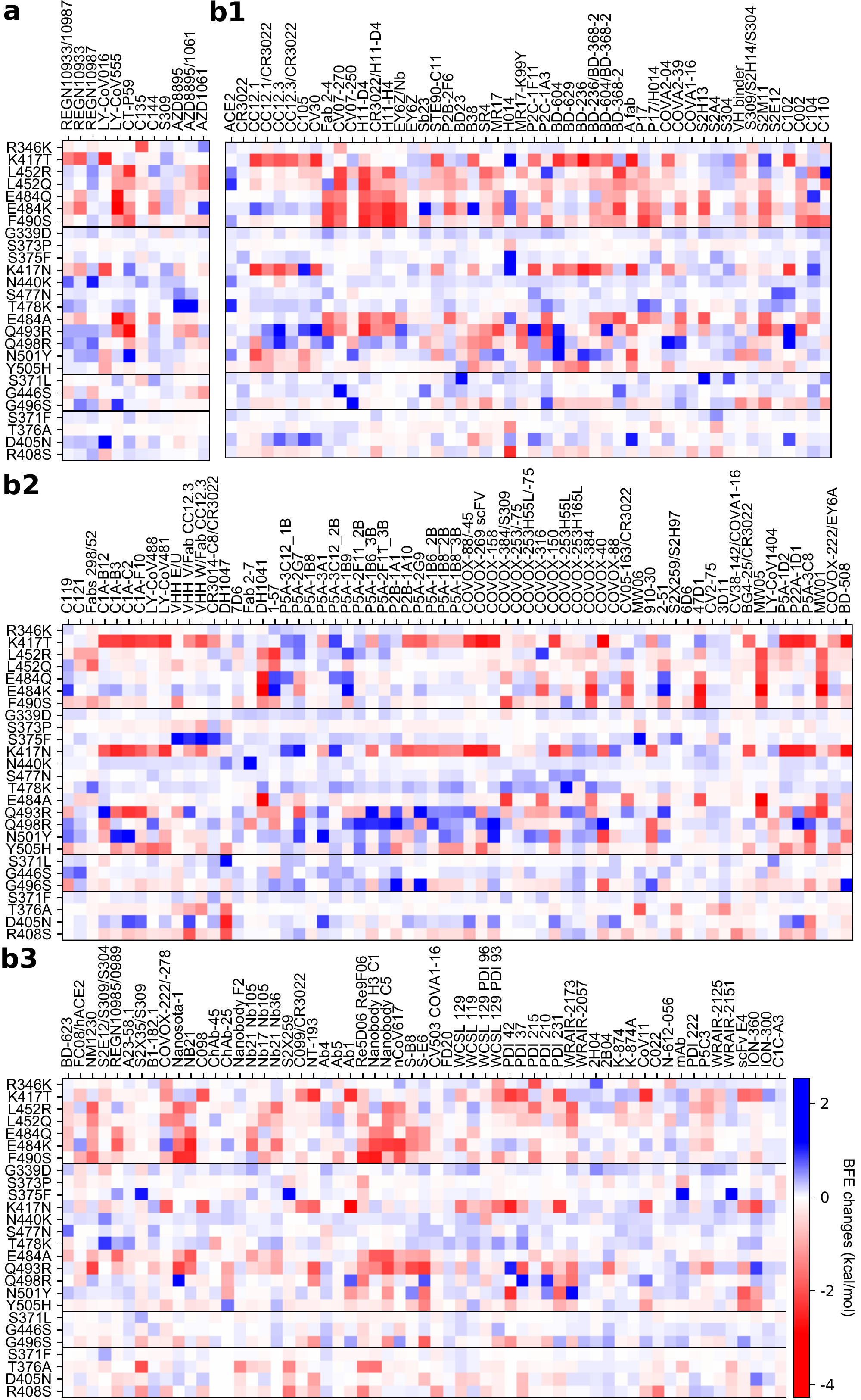}
\caption{Illustration of mutation-induced BFE changes of 185   antibody-RBD complexes and an ACE2-RBD complex. Positive changes strengthen the binding, while negative changes weaken the binding. 
		{\bf a} Heat map for 12 antibody-RBD complexes in various stages of drug development. Gray color stands for no predictions due to incomplete structures. 
		{\bf b1} Heat map for ACE2-RBD and antibody-RBD complexes. 
		{\bf b2} and {\bf b3} Heat map for antibody-RBD complexes. 
		The first 7 mutations are associated earlier SARS-CoV-2 variants. The next 12 mutations are shared among BA.1, BA.2, and BA.3 strains. The next three mutations are distinct to BA.1, and the final bunch of 4 mutations belong to BA.2.  
		}
	\label{fig:combine2_1}
\end{figure}

\begin{figure}[ht!]
	\centering
	\includegraphics[width = 0.8\textwidth]{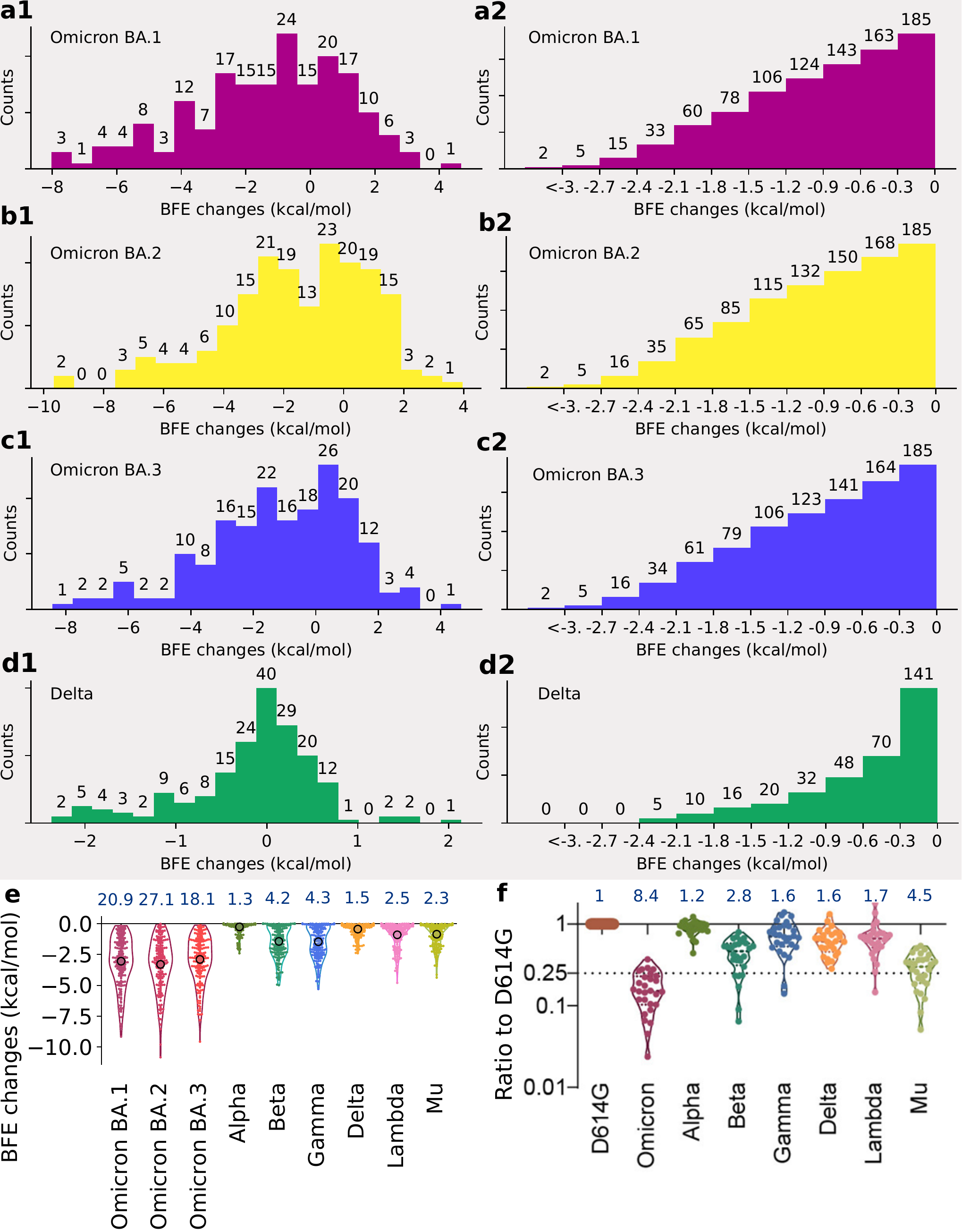}
\caption{Analysis of variant mutation-induced BFE changes of ACE2-RBD and 185  antibody-RBD complexes. {\bf a1, b1, c1,} and {\bf d1}  The distributions (counts) of accumulated BFE changes  induced by Omicron BA.1, BA.2, BA.3, and Delta   mutations respectively for 185 antibody-RBD complexes. For each case, there are more mutation-weakened complexes than mutation-strengthened  complexes.
{\bf a2, b2, c2,} and {\bf d2}  The numbers  of antibody-RBD complexes regarded as disrupted by  BA.1, BA.2, BA.3, and Delta  mutations respectively   under different thresholds ranging from 0 kcal/mol, -0.3 kcal/mol, to $<$-3 kcal/mol. 
{\bf e}  Accumulated negative BFE changes  induced by   BA.1, BA.2, BA3, Alpha, Beta, Delta, Gamma, Lambda, and Mu   mutations respectively for 185 antibody-RBD complexes. 
		For each variant, the number on the top is the fold of binding affinity reduction computed by $e^{-\rm BFE change_{average}}$, where ${\rm BFE change_{average}}$, marked by a circle,  is the mean value of  negative BFE changes for 185 antibody-RBD complexes.
		{\bf  f }   The comparison of neutralization activity against Omicron (BA.1), Alpha, Beta, Delta, Gamma, Lambda, and Mu  variants based on 28 convalescence sera \cite{zhang2022significant}. For each variant, the number on the top is the ratio of neutralization ED$_{50}$ compared to the reference strain D614G.  
}
	\label{fig:combine2_2}
\end{figure}
Figures~\ref{fig:combine2_1} {\bf a, b1}, and {\bf b2} depict the BFE changes of ACE2-RBD and 185 antibody-RBD complexes induced by the RBD mutations from SARS-CoV-2 variants. 
The first bunch of 7 mutations is associated with Alpha, Beta, Gamma, Delta, Lambda, and Mu. 
The second bunch of 12 mutations is shared among BA.1, BA.2, and BA.3. 
The next bunch of 3 mutations is associated with BA.1. The last bunch of 4 mutations belongs to BA2. Binding-strengthening mutations give rise to positive BFE changes, while binding-weakening mutations lead to negative BFE changes.
Obviously, shared Omicron mutations K417N, E484A, and Q493R are very disruptive to many antibodies.   BA.1 mutation G496S is also quite disruptive. BA.2 mutations T376A, D405N, and R408S may reduce the efficacy of many antibodies. 
Apparently, these complexes are significantly impacted by Omicron BA.1, BA.2, and BA.3 RBD mutations.  Overall, Figure \ref{fig:combine2_1} shows more negative BFE changes than positive ones, suggesting Omicron BA.1, BA.2, and BA3 mutations enable the breakthrough of current vaccines.

Statistical analysis of the BFE changes of 185 antibody-RBD complexes induced by  BA.1, BA.2, BA.3, and Delta RBD  mutations is presented in Figure \ref{fig:combine2_2} and analysis of Alpha, Beta, Gamma, Lambda, and Mu is presented in Figure~S2.    Accumulated BFE changes are provided in Figure \ref{fig:combine2_2} {\bf a1, b1}, and {\bf c1}. Obviously, all Omicron subvariants have more negative accumulated BFE changes than positive ones, showing their antibody resistance. Among them, BA.2's distribution is extended to a wider negative domain, showing its strongest antibody resistance. In contrast, Delta variant's statistics is given in Figure~\ref{fig:combine2_2} {\bf d1}, showing a smaller domain of distribution. 

As discussed earlier, it is difficult to obtain a variant's true vaccine-escape rate over  world's populations. However, a molecular-based comparative analysis can offer desirable information. Figures \ref{fig:combine2_2} {\bf a2, b2, c2,} and {\bf d2} depict the number of antibody-RBD complexes that is regarded as disrupted by   BA.1, BA.2, BA.3, and Delta  mutations, respectively,   under different thresholds ranging from 0 kcal/mol, -0.3 kcal/mol, to $<$-3 kcal/mol.  Previously, threshold  -0.3 kcal/mol was used to decide whether a mutation disrupts an antibody-RBD complex \cite{chen2022omicron}, which gives rise to  163, 168, and 164 disrupted antibody-RBD complexes, respectively for    BA.1, BA.2, and BA.3. The corresponding rates of potential vaccine breakthrough are 0.88, 0.91, and 0.89 for BA.1, BA.2, and BA.3, respectively. Therefore, BA.2 is slightly more antibody resistant than BA.1.  As a reference, the Delta variant may disrupt 70 out of 185 antibody-RBD complexes, suggesting a vaccine-breakthrough rate of 0.37. 
 
It is interesting to compare our analysis with experimental results
\cite{zhang2022significant}. In Figure \ref{fig:combine2_2} {\bf f}, the  sensitivity of 28 serum samples from COVID-19 convalescent patients
infected with an earlier SARS-CoV-2  strain (D614G) was tested against pseudotyped Omicron,  Alpha, Beta, Gamma, Delta, Lambda, and Mu  \cite{zhang2022significant}. The results indicate the Omicron (BA.1) and Delta variant have 8.4 and 1.6 fold reductions, respectively,  to the mean neutralization ED50 of these sera compared with the D614G reference strain.   Figure \ref{fig:combine2_2} {\bf e}  presents a comparison of accumulated negative BFE changes for variants Omicron BA.1, BA.2, BA.3, Alpha, Beta, Delta, Gamma, Lambda, and Mu. For each antibody-RBD complex, we only consider disruptive effects by setting positive BFE changes to zero and sum over RBD mutations (e.g., 15 mutations for Omicron BA.1 and 2 for Delta) to obtain the accumulated negative BFE change. As such, we have 185 accumulated negative BFE changes for each variant. We use the mean of these 185 values to computed the fold of affinity reduction, which can be compared for different variants against the original virus reported in Wuhan (${\rm BFE change_{average}}=0$). The RBD mutations of the Delta variant cause 1.5 fold reduction in the neutralization capability. In the same setting,  Omicron BA.1, BA.2, and BA.3 may lead to about 21, 27, and 18 fold increases in their vaccine-breakthrough capabilities. As such, BA.2 is about 30\% more capable to escape existing vaccines than BA.1 and 17 times more than the Delta variant.  Our prediction has a  correlation coefficient of 0.9 with the experiment.  With its highest infectivity and highest vaccine-escape potential, the Omicron BA.2 is set to take over the Omicron BA.1 in infecting the world population. 

\subsection{Antibody resistance}
\begin{figure}[ht!]
	\centering
	\includegraphics[width = 0.9\textwidth]{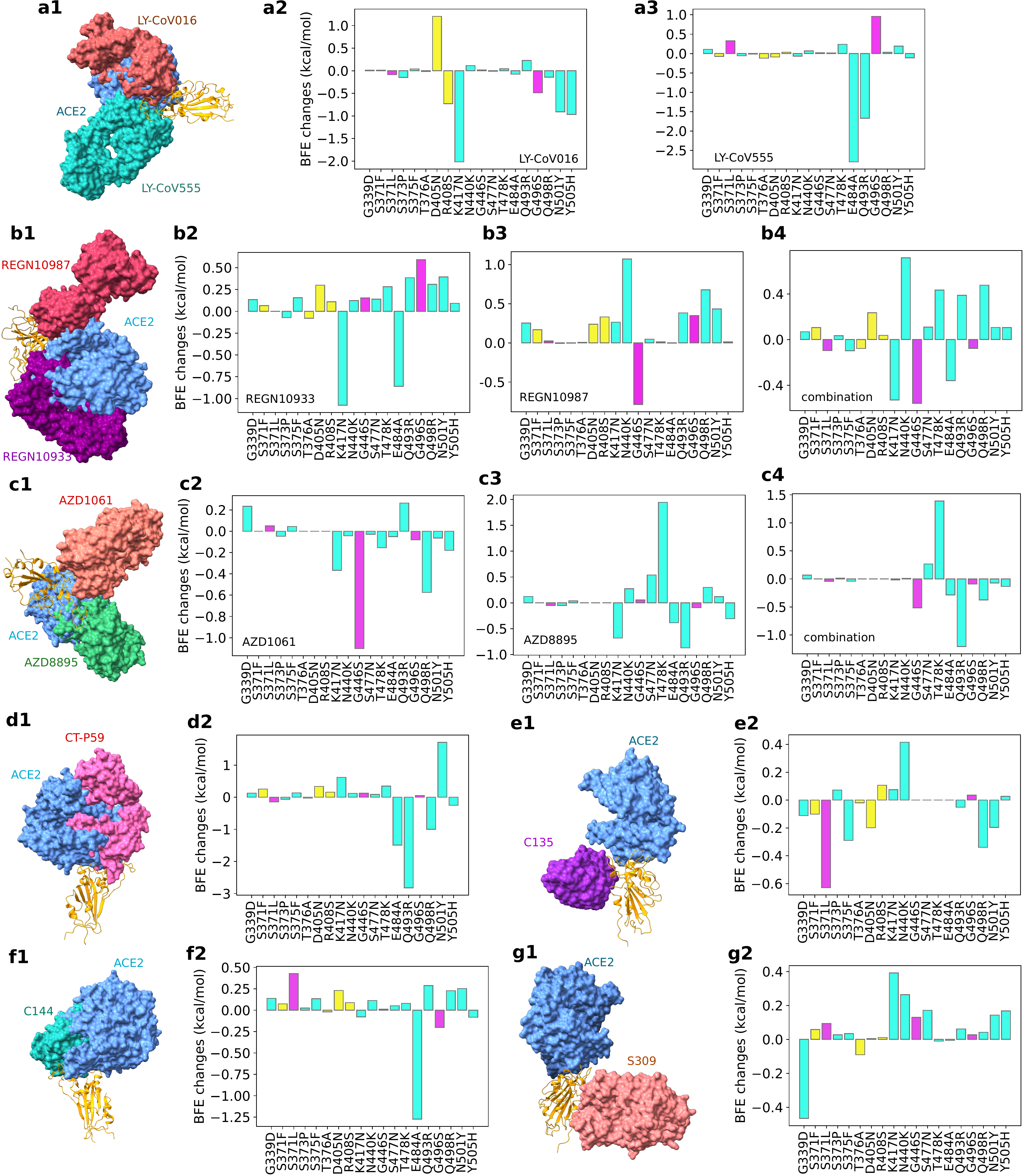}
\caption{Illustration of  Omicron BA.1 and BA.2 RBD mutational impacts on 
clinical mAbs. 
{\bf a1, b1, c1, d1, e1, f1} and {\bf g1} depict the 3D structures of  antibody-RBD complexes of    
Eli Lilly   LY-CoV555 (PDB ID: 7KMG \cite{jones2021neutralizing}) and LY-CoV016 (PDB ID: 7C01 \cite{shi2020human}), 
Regeneron    REGN10987 and REGN10933 (PDB ID: 6XDG \cite{hansen2020studies}),
AstraZeneca  AZD1061 and AZD8895 (PDB ID: 7L7E \cite{dong2021genetic}),
Celltrion   CT-P59 (aka Regdanvimab, PDB ID: 7CM4),
Rockefeller University   C135 (PDB ID: 7K8Z) and C144 (PDB ID: 7K90),
and 
GlaxoSmithKline  S309 (PDB ID: 6WPS), respectively. In all plots, the ACE2 structure is aligned as a reference.  
Omicron BA.1 and BA.2 RBD mutation-induced BFE changes (kcal/mol) are given in 
{\bf a2} and {\bf a3} for Eli Lilly mAbs, 
{b2, b3} and {\bf b4} for Regeneron  mAbs, 
{\bf c2, c3,} and {\bf c4} for AstraZeneca mAbs, 
{\bf d2} for Celltrion CT-P59, 
{\bf e2} and {\bf f2} for Rockefeller University mAbs,
and {\bf g2} for GlaxoSmithKline  S309, respectively. 
Cyan bars label the BFE changes induced by twelve RBD mutations shared by BA.1, BA.2, and BA.3 subvariants. 
Magenta bars mark  the BFE changes induced by three additional BA.1 RBD mutations.  Yellow bars denote  the BFE changes induced by  four additional BA.2 RBD mutations.
}
\label{fig:combine_mAbs}
\end{figure}

The design and discovery of mAbs are part of an important achievement in combating COVID-19. Unfortunately, like vaccines, mAbs are prone to viral mutations, particularly antibody-resistant ones. Early studies predicted that Omicron BA.1 would compromise the anti-COVID-19 mAbs developed by Eli Lilly, Regeneron,  AstraZeneca, Celltrion, and Rockefeller University \cite{chen2022omicron}. However,  Omicron BA.1's impact on   GlaxoSmithKline’s mAb, called sotrovimab, was predicted to be mild \cite{chen2022omicron}.  These predictions have been confirmed and the FDA has halted the use of Eli Lilly and  Regeneron's COVID-19 mAbs. Currently, GlaxoSmithKline's sotrovimab is the only antibody-drug authorized in the U.S. for the treatment of COVID-19 patients infected by the  Omicron variant. An important question is whether sotrovimab remains effective for the BA.2 subvariant that might drive a new wave of infections in the world population.

In this work, we further analyze the efficacy of these mAbs for BA.2 and BA.3. 
Our studies focus on  Omicron subvariants' RBD mutations, which appear to be optimized by the virus to evade host antibody protection and infect the host cell. Figure \ref{fig:combine_mAbs} provides a comprehensive analysis of the BFE changes of various antibody-RBD complexes induced by Omicron BA.1, BA.2, and BA.3. Since BA.3 subvariant's RBD mutations are the subsets of those of BA.1 and BA.2, we only present 19 unique RBD mutations.  Impacts of twelve shared RBD mutations are labeled with cyan, those of three additional BA.1 RBD mutations are marked with magenta, and those of four additional BA.2 RBD mutations are plotted in yellow.  
Figures \ref{fig:combine_mAbs} {\bf a1, b1, c1, d1, e1, f1} and {\bf g1} depict
3D antibody-RBD complexes for mAbs from 
Eli Lilly (LY-CoV016 and LY-CoV555),  
Regeneron (REGN10933, REGN10987, and REGN10933/10987), 
AstraZeneca (AZD1061 and AZD8895),  
Celltrion (CT-P59),  
Rockefeller University (C135, C144), and  
GlaxoSmithKline (S309), respectively.  The ACE2 is included in these plots as a reference. 

Figures \ref{fig:combine_mAbs} {\bf  a2} and {\bf a3} show that LY-CoV016 is disrupted by shared mutation K417N and LY-CoV555 is weakened by shared mutations E484A and Q493R.  Additional mutations from BA.2 may not significantly affect Eli Lilly mAbs. However, if BA.2 become dominant,  Eli Lilly mAbs would still be ineffective. 

The impacts of BA.1 and BA.2 mutations on Regeneron's mAbs are illustrated in Figures \ref{fig:combine_mAbs} {\bf  b2, b3} and {\bf b4}.  REGN10933 is undermined by shared mutations N417K and E484A.  REGN10987 is disrupted by BA.1 mutation G446S. The antibody cocktail is undermined by shared Omicron mutations as well, which implies Regeneron's mAbs would still be compromised should Omicron BA.2 become a dominant SRAS-CoV-2 subvariant. 

BA.1 and BA.2's impacts on AstraZeneca's AZD1061 and AZD8895 are demonstrated in  Figures \ref{fig:combine_mAbs} {\bf  c2, c3} and {\bf c4}. It is noticed that BA.1 mutation G446S has a disruptive effect on AZD1061. AZD8895 is weakened by two shared mutations. The AZD1061-AZD8895 combination is also disrupted by shared mutation Q493R. Therefore,   the efficacy of AstraZeneca's mAbs would be reduced should BA.2 prevail in world populations.

As shown in Figure  \ref{fig:combine_mAbs} {\bf d2}, Celltrion's mAb CT-P59 is prone to shared mutations Q493R and E484A. BA.2 mutations may not bring additional destruction. However, the shared mutations pose a threat to Celltrion's mAb, which implies its efficacy would not restore should BA.2 prevail.    

Figures  \ref{fig:combine_mAbs} {\bf e2} and {\bf f2} present BA.1 and BA.2's mutational impacts on Rockefeller University's mAbs. C135 is mainly disrupted by Omicron BA.1 and its C144 is made ineffective by shared mutation E484A. Therefore,  C135 might become effective if BA.2 dominates. 

Finally, we plot mutational impacts on antibody S309's binding with RBD in Figure  \ref{fig:combine_mAbs} {\bf g2}. Antibody S309  is the parent antibody for Sotrovimab developed by GlaxoSmithKline and Vir Biotechnology, Inc. It is seen from the figure that there is only one disruptive BFE change of -0.47kcal/mol and the rest of the BFE changes are mostly positive.  The BA.2 mutations have little effect on S309. Therefore, we expect a mild effect from Omicron BA.1 and BA.2 on  sotrovimab. 
 
It is interesting to understand why S309 is the only antibody that is not significantly affected by Omicron variants. Figure   \ref{fig:combine_mAbs}
show that all mAbs that compete with the human ACE2 for the receptor-binding motif (RBM) are seriously compromised by Omicron subvariants because most of the RBD mutations locate at the RBM. A possible reason is that Omicron subvariants had optimized RBD mutations at the RBM to strengthen the viral infectivity and evade the direct neutralization antibodies. Consequently, all mAbs that target RBM are seriously compromised by Omicron subvariants. Figures  \ref{fig:combine_mAbs} {\bf  e1} and {\bf g1} show that antibodies C135 and  S309 do not directly compete with ACE2 for the RBM.   However,  C135  is still very close to the RBM and significantly weakened by some Omicron mutations. In contrast, S309 is further away from the RBM and escapes from Omicron's RBD mutations.

\section{Materials and Methods}
%{\bf Machine learning model.} 
The deep learning model is designed for predicting mutation-induced BFE changes of the binding between protein-protein interactions. A series of three steps consist of training data preparation, feature generations, and deep neural network training and prediction (see Figure~S2). Here, we briefly discuss each step and leave more details in Supporting Information. Readers are also suggested literature \cite{cang2017analysis,chen2020mutations,chen2021revealing} for more details.

Firstly, the training data is prepared to comprise experimental BFE changes and next-generation sequencing data. SKEMPI 2.0 \cite{jankauskaite2019skempi} is the fundamental BFE change dataset. Additionally, SARS-CoV-2 related datasets are the mutational scanning data of the ACE2-RBD complex \cite{chan2020engineering,starr2020deep,linsky2020novo} and the CTC-445.2-RBD complex \cite{linsky2020novo}. Next is to prepare the features. It is required a variety of biochemical, biophysical, and mathematics features from PPI complex structures, such as surface areas, partial charges, van der Waals interaction, Coulomb interactions, pH values, electrostatics, persistent homology, graph theory, etc.\cite{chen2020mutations,wang2020topology} A detailed list and description of these features are provided in Supporting Information. In the following, the key idea of the element-specific and site-specific persistent homology is illustrated briefly. As the persistent homology \cite{zomorodian2005computing,edelsbrunner2008persistent} introduced as a useful tool for data analysis for scientific and engineering applications, it is further applied to molecular studies \cite{cang2017analysis, cang2018representability}. For 3D structures, atoms are modeled as vertices in a point cloud. Then edges, faces, etc. can be constructed as simplices $\sigma$ which form simplicial complexes $X$. Groups $C_k(X)$, $k=0,1,2,3$ are sets of all chains of $k$th dimension, which is defined as a finite sum of simplices as $\sum_{i}\alpha_i\sigma^k_i$ with coefficients $\alpha_i$. The boundary operator $\partial_k$ therefore, maps $C_k(X)\!\rightarrow\!C_{k-1}(X)$ as
\begin{equation}
\partial_k \sigma^k = \sum^{k}_{i=0} (-1)^i[v_0, \cdots, \hat{v}_i, \cdots, v_k ],
\end{equation}
where $\sigma^k=\{v_0,\cdots,v_k\}$ and $[v_0, \cdots ,\hat{v}_i, \cdots, v_k]$ is a $(k\!-\!1)$-simplex excluding $v_i$ with $\partial_{k-1}\partial_k= 0$. The chain complex is given as
\begin{equation}
\cdots \stackrel{\partial_{k+1}}\longrightarrow C_k(X) \stackrel{\partial_k}\longrightarrow C_{k-1}(X) \stackrel{\partial_{k-1}}\longrightarrow \cdots \stackrel{\partial_2} \longrightarrow C_{1}(X) \stackrel{\partial_{1}}\longrightarrow C_0(X) \stackrel{\partial_0} \longrightarrow 0.
\end{equation}
The $k$-th homology group $H_k$ is defined by $H_k = Z_k / B_k$ where $Z_k={\rm ker} ~\partial_k=\{c\in C_k \mid \partial_k c=0\}$ and $B_k={\rm im} ~\partial_{k+1}= \{ \partial_{k+1} c \mid c\in C_{k+1}\}$. Thus, the Betti numbers can be defined by the ranks of $k$-th homology group $H_k$. Persistent homology can be devised to track Betti numbers through a filtration where $\beta_0$ describes the number of connected components, $\beta_1$ provides the number of loops, and $\beta_2$ is the number of cavities. Therefore, using persistent homology, the atoms of 3D structures are grouped according to their elements, as well as the atoms from the binding site of antibodies and antibodies. The interactions and their impacts on PPI complex bindings are characterized by the topological invariants, which are further implemented for machine learning training.

Lastly, a deep learning algorithm, artificial/deep neural networks (ANNs or DNNs), is used to tackle the features with datasets for training and predictions\cite{chen2021revealing}. A trained model is available at \href{https://github.com/WeilabMSU/TopNetmAb}{TopNetmAb}, a SARS-CoV-2-specific model, whose early model was integrating convolutional neural networks (CNNs) with gradient boosting trees (GBTs) and was trained only on the SKEMPI 2.0 dataset with a high accuracy \cite{wang2020topology}.

Recent work with predictions from TopNetmAb \cite{chen2021prediction,chen2021revealing,wang2021emerging} is highly consistent with experimental results. One should notice it is important with the help of the aforementioned deep mutational datasets related to SARS-CoV-2. The Pearson correlation of our predictions for the binding of CTC-445.2 and RBD with experimental data is 0.7\cite{chen2021revealing,linsky2020novo}. Meanwhile, a Pearson correlation of 0.8 is observed of the predictions of clinical trial antibodies against SARS-CoV-2 induced by emerging mutations in the same work\cite{chen2021revealing} compared to the natural log of experimental escape fractions\cite{starr2021prospective}. Moreover, the prediction of single mutations L452R and N501Y for the ACE2-RBD complex have a perfect consistency with experimental luciferase data \cite{chen2021revealing,deng2021transmission}. More detailed validations are in Supporting Information.

\section{Conclusion}

The Omicron variant has three subvariants BA.1, BA.2, and BA3. The Omicron BA.1 has surprised the scientific community by its large number of mutations, particularly those on the spike (S) protein receptor-binding domain (RBD), which enable its unusual infectivity and high ability to evade antibody protections induced by viral infection and vaccination. Viral RBD interacts with host angiotensin-converting enzyme 2 (ACE2) to initiate cell entry and infection and is a major target for vaccines and monoclonal antibodies (mAbs). Omicron BA.1 exploits its 15 RBD mutations to strengthen its infectivity and disrupt mAbs generated by prior viral infection or vaccination. Omicron BA.2 and BA.3 share 12 RBD mutations with BA.1 but differ by 4 and 3 RBD mutations, respectively, suggesting potentially serious threats to human health. However, no experimental result has been reported for Omicron BA.2 and BA.3, although BA.2 is found to be able to alarmingly reinfect patients originally infected by Omicron BA.1 \cite{timesofisrael}. In this work, we present deep learning predictions of BA.2's and BA.3's potential to become another dominating variant. Based on an intensively tested deep learning model trained with tens of thousands of experimental data, we investigate Omicron BA.2's and BA.3's RBD mutational impacts on the RBD-ACE2 binding complex to understand their infectivity and a library of 185 antibodies to shed light on their threats to vaccines and existing mAbs.   We unveil that BA.2 is about 1.5 and 4.2 times as contagious as BA.1 and Delta, respectively. It is also 30\% and 17-fold more capable than BA.1 and Delta, respectively, to escape current vaccines. It is predicted to undermine most existing mAbs,   except for sotrovimab developed by GlaxoSmithKline. We forecast Omicron BA.2 will become another prevailing variant by infecting populations with or without antibody protection.   

\section*{Data and model availability}
The structural information of 185 antibody-RBD complexes with their corresponding PDB IDs and the results of BFE changes of PPI complexes induced by mutations can be found in Section S2 of the Supporting Information. The TopNetTree model is available at \href{https://github.com/WeilabMSU/TopNetmAb}{TopNetmAb}. 
The detailed methods can be found in the Supporting Information S3 and S4. The validation of our predictions with experimental data can be located in Supporting Information S5.

\section*{Supporting information}
The supporting information is available for 
\begin{enumerate}
	\item[S1] Supplementary figures: analysis of variant mutation-induced BFE changes for Alpha, Beta, Gamma, Lambda, and Mu variants (the extension of Figure~\ref{fig:combine2_2}). 
	\item[S2] Supplementary data: The Supplementary\_Data.zip contains two files: the BFE changes of antibodies disrupted by Omicron subvariant mutations and the list of antibodies with corresponding PDB IDs
	\item[S3] Supplementary feature generation methods
	\item[S4] Supplementary machine learning methods
	\item[S5] Supplementary validation: validations of our machine learning predictions with experimental data
\end{enumerate}

\section*{Acknowledgment}
This work was supported in part by NIH grant  GM126189, NSF grants DMS-2052983,  DMS-1761320, and IIS-1900473,  NASA grant 80NSSC21M0023,  Michigan Economic Development Corporation, MSU Foundation,  Bristol-Myers Squibb 65109, and Pfizer.

% \bibliographystyle{abbrv}
%\bibliographystyle{unsrt}
%% \bibliographystyle{custom}
%\bibliography{refs}

\begin{thebibliography}{10}
	
	\bibitem{chen2022omicron}
	Jiahui Chen, Rui Wang, Nancy~Benovich Gilby, and Guo-Wei Wei.
	\newblock Omicron variant (b. 1.1. 529): Infectivity, vaccine breakthrough, and
	antibody resistance.
	\newblock {\em Journal of chemical information and modeling}, 2022.
	
	\bibitem{shuai2022attenuated}
	Huiping Shuai, Jasper Fuk-Woo Chan, Bingjie Hu, Yue Chai, Terrence Tsz-Tai
	Yuen, Feifei Yin, Xiner Huang, Chaemin Yoon, Jing-Chu Hu, Huan Liu, et~al.
	\newblock Attenuated replication and pathogenicity of sars-cov-2 b. 1.1. 529
	omicron.
	\newblock {\em Nature}, pages 1--1, 2022.
	
	\bibitem{hong2022molecular}
	Qin Hong, Wenyu Han, Jiawei Li, Shiqi Xu, Yifan Wang, Zuyang Li, Yanxing Wang,
	Chao Zhang, Zhong Huang, and Yao Cong.
	\newblock Molecular basis of sars-cov-2 omicron variant receptor engagement and
	antibody evasion and neutralization.
	\newblock {\em bioRxiv}, 2022.
	
	\bibitem{cele2021omicron}
	Sandile Cele, Laurelle Jackson, David~S Khoury, Khadija Khan, Thandeka
	Moyo-Gwete, Houriiyah Tegally, James~Emmanuel San, Deborah Cromer, Cathrine
	Scheepers, Daniel~G Amoako, et~al.
	\newblock Omicron extensively but incompletely escapes pfizer bnt162b2
	neutralization.
	\newblock {\em Nature}, pages 1--5, 2021.
	
	\bibitem{zhang2022significant}
	Li~Zhang, Qianqian Li, Ziteng Liang, Tao Li, Shuo Liu, Qianqian Cui, Jianhui
	Nie, Qian Wu, Xiaowang Qu, Weijin Huang, et~al.
	\newblock The significant immune escape of pseudotyped sars-cov-2 variant
	omicron.
	\newblock {\em Emerging microbes \& infections}, 11(1):1--5, 2022.
	
	\bibitem{liu2021striking}
	Lihong Liu, Sho Iketani, Yicheng Guo, Jasper~FW Chan, Maple Wang, Liyuan Liu,
	Yang Luo, Hin Chu, Yiming Huang, Manoj~S Nair, et~al.
	\newblock Striking antibody evasion manifested by the omicron variant of
	sars-cov-2.
	\newblock {\em Nature}, pages 1--8, 2021.
	
	\bibitem{lu2021neutralization}
	Lu~Lu, Bobo Wing-Yee Mok, Linlei Chen, Jacky Man-Chun Chan, Owen Tak-Yin Tsang,
	Bosco Hoi-Shiu Lam, Vivien Wai-Man Chuang, Allen Wing-Ho Chu, Wan-Mui Chan,
	Jonathan~Daniel Ip, et~al.
	\newblock Neutralization of sars-cov-2 omicron variant by sera from bnt162b2 or
	coronavac vaccine recipients.
	\newblock {\em Clin Infect Dis, doi:10.1093/cid/ciab1041}, 2021.
	
	\bibitem{hoffmann2021omicron}
	Markus Hoffmann, Nadine Kr{\"u}ger, Sebastian Schulz, Anne Cossmann, Cheila
	Rocha, Amy Kempf, Inga Nehlmeier, Luise Graichen, Anna-Sophie Moldenhauer,
	Martin~S Winkler, et~al.
	\newblock The omicron variant is highly resistant against antibody-mediated
	neutralization--implications for control of the covid-19 pandemic.
	\newblock {\em Cell}, 2021.
	
	\bibitem{desingu2022emergence}
	Perumal~Arumugam Desingu, K~Nagarajan, and Kuldeep Dhama.
	\newblock Emergence of omicron third lineage ba. 3 and its importance.
	\newblock {\em Journal of Medical Virology}, 2022.
	
	\bibitem{walls2020structure}
	Alexandra~C Walls, Young-Jun Park, M~Alejandra Tortorici, Abigail Wall,
	Andrew~T McGuire, and David Veesler.
	\newblock Structure, function, and antigenicity of the {SARS-CoV-2} spike
	glycoprotein.
	\newblock {\em Cell}, 2020.
	
	\bibitem{lyngse2022transmission}
	Frederik~Plesner Lyngse, Carsten~Thure Kirkeby, Matthew Denwood, Lasse~Engbo
	Christiansen, K{\aa}re M{\o}lbak, Camilla~Holten M{\o}ller, Robert~Leo Skov,
	Tyra~Grove Krause, Morten Rasmussen, Raphael~Niklaus Sieber, et~al.
	\newblock Transmission of sars-cov-2 omicron voc subvariants ba. 1 and ba. 2:
	Evidence from danish households.
	\newblock {\em medRxiv}, 2022.
	
	\bibitem{timesofisrael}
	{BA2reinfection}.
	\newblock
	\url{https://www.timesofisrael.com/several-cases-of-omicron-reinfection-said-detected-in-israel-with-new-ba2-strain/}.
	
	\bibitem{world2021enhancing}
	World~Health Organization et~al.
	\newblock Enhancing readiness for omicron (b. 1.1. 529): technical brief and
	priority actions for member states, 2021.
	
	\bibitem{li2005bats}
	Wendong Li, Zhengli Shi, Meng Yu, Wuze Ren, Craig Smith, Jonathan~H Epstein,
	Hanzhong Wang, Gary Crameri, Zhihong Hu, Huajun Zhang, et~al.
	\newblock Bats are natural reservoirs of {SARS}-like coronaviruses.
	\newblock {\em Science}, 310(5748):676--679, 2005.
	
	\bibitem{hoffmann2020sars}
	Markus Hoffmann, Hannah Kleine-Weber, Simon Schroeder, Nadine Kr{\"u}ger, Tanja
	Herrler, Sandra Erichsen, Tobias~S Schiergens, Georg Herrler, Nai-Huei Wu,
	Andreas Nitsche, et~al.
	\newblock {SARS-CoV-2} cell entry depends on {ACE2} and {TMPRSS2} and is
	blocked by a clinically proven protease inhibitor.
	\newblock {\em Cell}, 181(2):271--280, 2020.
	
	\bibitem{chen2020mutations}
	Jiahui Chen, Rui Wang, Menglun Wang, and Guo-Wei Wei.
	\newblock Mutations strengthened {SARS-CoV-2} infectivity.
	\newblock {\em Journal of molecular biology}, 432(19):5212--5226, 2020.
	
	\bibitem{wang2021vaccine}
	Rui Wang, Jiahui Chen, Kaifu Gao, and Guo-Wei Wei.
	\newblock Vaccine-escape and fast-growing mutations in the {United Kingdom, the
		United States, Singapore, Spain, India}, and other {COVID}-19-devastated
	countries.
	\newblock {\em Genomics}, 113(4):2158--2170, 2021.
	
	\bibitem{wang2020human}
	Chunyan Wang, Wentao Li, Dubravka Drabek, Nisreen~MA Okba, Rien van Haperen,
	Albert~DME Osterhaus, Frank~JM van Kuppeveld, Bart~L Haagmans, Frank
	Grosveld, and Berend-Jan Bosch.
	\newblock A human monoclonal antibody blocking {SARS-CoV-2} infection.
	\newblock {\em Nature communications}, 11(1):1--6, 2020.
	
	\bibitem{yu2020receptor}
	Fei Yu, Rong Xiang, Xiaoqian Deng, Lili Wang, Zhengsen Yu, Shijun Tian, Ruiying
	Liang, Yanbai Li, Tianlei Ying, and Shibo Jiang.
	\newblock Receptor-binding domain-specific human neutralizing monoclonal
	antibodies against {SARS-CoV} and {SARS-CoV-2}.
	\newblock {\em Signal Transduction and Targeted Therapy}, 5(1):1--12, 2020.
	
	\bibitem{li2021impact}
	Cheng Li, Xiaolong Tian, Xiaodong Jia, Jinkai Wan, Lu~Lu, Shibo Jiang, Fei Lan,
	Yinying Lu, Yanling Wu, and Tianlei Ying.
	\newblock The impact of receptor-binding domain natural mutations on antibody
	recognition of {SARS-CoV-2}.
	\newblock {\em Signal Transduction and Targeted Therapy}, 6(1):1--3, 2021.
	
	\bibitem{zhu2021cryo}
	X~Zhu, D~Mannar, JW~Saville, SS~Srivastava, AM~Berezuk, KS~Tuttle, and
	S~Subramaniam.
	\newblock Cryo-em structure of sars-cov-2 omicron spike protein in complex with
	human ace2 (focused refinement of rbd and ace2), 2021.
	
	\bibitem{chen2021prediction}
	Jiahui Chen, Kaifu Gao, Rui Wang, and Guo-Wei Wei.
	\newblock Prediction and mitigation of mutation threats to {COVID-19} vaccines
	and antibody therapies.
	\newblock {\em Chemical Science}, 12(20):6929--6948, 2021.
	
	\bibitem{jones2021neutralizing}
	Bryan~E Jones, Patricia~L Brown-Augsburger, Kizzmekia~S Corbett, Kathryn
	Westendorf, Julian Davies, Thomas~P Cujec, Christopher~M Wiethoff, Jamie~L
	Blackbourne, Beverly~A Heinz, Denisa Foster, et~al.
	\newblock The neutralizing antibody, ly-cov555, protects against sars-cov-2
	infection in nonhuman primates.
	\newblock {\em Science translational medicine}, 13(593), 2021.
	
	\bibitem{shi2020human}
	Rui Shi, Chao Shan, Xiaomin Duan, Zhihai Chen, Peipei Liu, Jinwen Song, Tao
	Song, Xiaoshan Bi, Chao Han, Lianao Wu, et~al.
	\newblock A human neutralizing antibody targets the receptor binding site of
	{SARS-CoV-2}.
	\newblock {\em Nature}, pages 1--8, 2020.
	
	\bibitem{hansen2020studies}
	Johanna Hansen, Alina Baum, Kristen~E Pascal, Vincenzo Russo, Stephanie
	Giordano, Elzbieta Wloga, Benjamin~O Fulton, Ying Yan, Katrina Koon, Krunal
	Patel, et~al.
	\newblock Studies in humanized mice and convalescent humans yield a
	{SARS-CoV-2} antibody cocktail.
	\newblock {\em Science}, 369(6506):1010--1014, 2020.
	
	\bibitem{dong2021genetic}
	Jinhui Dong, Seth~J Zost, Allison~J Greaney, Tyler~N Starr, Adam~S Dingens,
	Elaine~C Chen, Rita~E Chen, James~Brett Case, Rachel~E Sutton, Pavlo Gilchuk,
	et~al.
	\newblock Genetic and structural basis for sars-cov-2 variant neutralization by
	a two-antibody cocktail.
	\newblock {\em Nature Microbiology}, 6(10):1233--1244, 2021.
	
	\bibitem{cang2017analysis}
	Zixuan Cang and Guo-Wei Wei.
	\newblock Analysis and prediction of protein folding energy changes upon
	mutation by element specific persistent homology.
	\newblock {\em Bioinformatics}, 33(22):3549--3557, 2017.
	
	\bibitem{chen2021revealing}
	Jiahui Chen, Kaifu Gao, Rui Wang, and Guo-Wei Wei.
	\newblock Revealing the threat of emerging {SARS-CoV-2} mutations to antibody
	therapies.
	\newblock {\em Journal of Molecular Biology}, 433(7744), 2021.
	
	\bibitem{jankauskaite2019skempi}
	Justina Jankauskait{\.e}, Brian Jim{\'e}nez-Garc{\'\i}a, Justas Dapk{\=u}nas,
	Juan Fern{\'a}ndez-Recio, and Iain~H Moal.
	\newblock {SKEMPI} 2.0: an updated benchmark of changes in protein--protein
	binding energy, kinetics and thermodynamics upon mutation.
	\newblock {\em Bioinformatics}, 35(3):462--469, 2019.
	
	\bibitem{chan2020engineering}
	Kui~K Chan, Danielle Dorosky, Preeti Sharma, Shawn~A Abbasi, John~M Dye,
	David~M Kranz, Andrew~S Herbert, and Erik Procko.
	\newblock Engineering human {ACE2} to optimize binding to the spike protein of
	{SARS} coronavirus 2.
	\newblock {\em Science}, 369(6508):1261--1265, 2020.
	
	\bibitem{starr2020deep}
	Tyler~N Starr, Allison~J Greaney, Sarah~K Hilton, Daniel Ellis, Katharine~HD
	Crawford, Adam~S Dingens, Mary~Jane Navarro, John~E Bowen, M~Alejandra
	Tortorici, Alexandra~C Walls, et~al.
	\newblock Deep mutational scanning of {SARS-CoV-2} receptor binding domain
	reveals constraints on folding and {ACE2} binding.
	\newblock {\em Cell}, 182(5):1295--1310, 2020.
	
	\bibitem{linsky2020novo}
	Thomas~W Linsky, Renan Vergara, Nuria Codina, Jorgen~W Nelson, Matthew~J
	Walker, Wen Su, Christopher~O Barnes, Tien-Ying Hsiang, Katharina
	Esser-Nobis, Kevin Yu, et~al.
	\newblock De novo design of potent and resilient {hACE2} decoys to neutralize
	{SARS-CoV-2}.
	\newblock {\em Science}, 370(6521):1208--1214, 2020.
	
	\bibitem{wang2020topology}
	Menglun Wang, Zixuan Cang, and Guo-Wei Wei.
	\newblock A topology-based network tree for the prediction of protein--protein
	binding affinity changes following mutation.
	\newblock {\em Nature Machine Intelligence}, 2(2):116--123, 2020.
	
	\bibitem{zomorodian2005computing}
	Afra Zomorodian and Gunnar Carlsson.
	\newblock Computing persistent homology.
	\newblock {\em Discrete \& Computational Geometry}, 33(2):249--274, 2005.
	
	\bibitem{edelsbrunner2008persistent}
	Herbert Edelsbrunner, John Harer, et~al.
	\newblock Persistent homology-a survey.
	\newblock {\em Contemporary mathematics}, 453:257--282, 2008.
	
	\bibitem{cang2018representability}
	Zixuan Cang, Lin Mu, and Guo-Wei Wei.
	\newblock Representability of algebraic topology for biomolecules in machine
	learning based scoring and virtual screening.
	\newblock {\em PLoS computational biology}, 14(1):e1005929, 2018.
	
	\bibitem{wang2021emerging}
	Rui Wang, Jiahui Chen, Yuta Hozumi, Changchuan Yin, and Guo-Wei Wei.
	\newblock Emerging vaccine-breakthrough {SARS-CoV-2} variants.
	\newblock {\em ACS Infect. Dis.}, 2021.
	
	\bibitem{starr2021prospective}
	Tyler~N Starr, Allison~J Greaney, Amin Addetia, William~W Hannon, Manish~C
	Choudhary, Adam~S Dingens, Jonathan~Z Li, and Jesse~D Bloom.
	\newblock Prospective mapping of viral mutations that escape antibodies used to
	treat {COVID-19}.
	\newblock {\em Science}, 371(6531):850--854, 2021.
	
	\bibitem{deng2021transmission}
	Xianding Deng, Miguel~A Garcia-Knight, Mir~M Khalid, Venice Servellita, Candace
	Wang, Mary~Kate Morris, Alicia Sotomayor-Gonz{\'a}lez, Dustin~R Glasner,
	Kevin~R Reyes, Amelia~S Gliwa, et~al.
	\newblock Transmission, infectivity, and antibody neutralization of an emerging
	{SARS-CoV-2} variant in {C}alifornia carrying a {L452R} spike protein
	mutation.
	\newblock {\em MedRxiv}, 2021.
	
\end{thebibliography}

\begin{thebibliography}{10}
	
	\bibitem{jankauskaite2019skempi}
	Justina Jankauskait{\.e}, Brian Jim{\'e}nez-Garc{\'\i}a, Justas Dapk{\=u}nas,
	Juan Fern{\'a}ndez-Recio, and Iain~H Moal.
	\newblock {SKEMPI} 2.0: an updated benchmark of changes in protein--protein
	binding energy, kinetics and thermodynamics upon mutation.
	\newblock {\em Bioinformatics}, 35(3):462--469, 2019.
	
	\bibitem{procko2020sequence}
	Erik Procko.
	\newblock The sequence of human ace2 is suboptimal for binding the s spike
	protein of sars coronavirus 2.
	\newblock {\em BioRxiv}, 2020.
	
	\bibitem{starr2020deep}
	Tyler~N Starr, Allison~J Greaney, Sarah~K Hilton, Daniel Ellis, Katharine~HD
	Crawford, Adam~S Dingens, Mary~Jane Navarro, John~E Bowen, M~Alejandra
	Tortorici, Alexandra~C Walls, et~al.
	\newblock Deep mutational scanning of {SARS-CoV-2} receptor binding domain
	reveals constraints on folding and {ACE2} binding.
	\newblock {\em Cell}, 182(5):1295--1310, 2020.
	
	\bibitem{linsky2020novo}
	Thomas~W Linsky, Renan Vergara, Nuria Codina, Jorgen~W Nelson, Matthew~J
	Walker, Wen Su, Christopher~O Barnes, Tien-Ying Hsiang, Katharina
	Esser-Nobis, Kevin Yu, et~al.
	\newblock De novo design of potent and resilient {hACE2} decoys to neutralize
	{SARS-CoV-2}.
	\newblock {\em Science}, 370(6521):1208--1214, 2020.
	
	\bibitem{chen2021revealing}
	Jiahui Chen, Kaifu Gao, Rui Wang, and Guo-Wei Wei.
	\newblock Revealing the threat of emerging {SARS-CoV-2} mutations to antibody
	therapies.
	\newblock {\em Journal of Molecular Biology}, 433(7744), 2021.
	
	\bibitem{carlsson2009topology}
	Gunnar Carlsson.
	\newblock Topology and data.
	\newblock {\em Bulletin of the American Mathematical Society}, 46(2):255--308,
	2009.
	
	\bibitem{edelsbrunner2000topological}
	Herbert Edelsbrunner, David Letscher, and Afra Zomorodian.
	\newblock Topological persistence and simplification.
	\newblock In {\em Proceedings 41st annual symposium on foundations of computer
		science}, pages 454--463. IEEE, 2000.
	
	\bibitem{xia2014persistent}
	Kelin Xia and Guo-Wei Wei.
	\newblock Persistent homology analysis of protein structure, flexibility, and
	folding.
	\newblock {\em International journal for numerical methods in biomedical
		engineering}, 30(8):814--844, 2014.
	
	\bibitem{chen2021prediction}
	Jiahui Chen, Kaifu Gao, Rui Wang, and Guo-Wei Wei.
	\newblock Prediction and mitigation of mutation threats to {COVID-19} vaccines
	and antibody therapies.
	\newblock {\em Chemical Science}, 12(20):6929--6948, 2021.
	
	\bibitem{chen2020mutations}
	Jiahui Chen, Rui Wang, Menglun Wang, and Guo-Wei Wei.
	\newblock Mutations strengthened {SARS-CoV-2} infectivity.
	\newblock {\em Journal of molecular biology}, 432(19):5212--5226, 2020.
	
	\bibitem{wang2020mutations}
	Rui Wang, Yuta Hozumi, Changchuan Yin, and Guo-Wei Wei.
	\newblock Mutations on {COVID-19} diagnostic targets.
	\newblock {\em Genomics}, 112(6):5204--5213, 2020.
	
	\bibitem{wang2020topology}
	Menglun Wang, Zixuan Cang, and Guo-Wei Wei.
	\newblock A topology-based network tree for the prediction of protein--protein
	binding affinity changes following mutation.
	\newblock {\em Nature Machine Intelligence}, 2(2):116--123, 2020.
	
	\bibitem{bas2008very}
	Delphine~C Bas, David~M Rogers, and Jan~H Jensen.
	\newblock Very fast prediction and rationalization of pka values for
	protein--ligand complexes.
	\newblock {\em Proteins: Structure, Function, and Bioinformatics},
	73(3):765--783, 2008.
	
	\bibitem{altschul1997gapped}
	Stephen~F Altschul, Thomas~L Madden, Alejandro~A Sch{\"a}ffer, Jinghui Zhang,
	Zheng Zhang, Webb Miller, and David~J Lipman.
	\newblock Gapped blast and psi-blast: a new generation of protein database
	search programs.
	\newblock {\em Nucleic acids research}, 25(17):3389--3402, 1997.
	
	\bibitem{yang2017spider2}
	Yuedong Yang, Rhys Heffernan, Kuldip Paliwal, James Lyons, Abdollah Dehzangi,
	Alok Sharma, Jihua Wang, Abdul Sattar, and Yaoqi Zhou.
	\newblock Spider2: A package to predict secondary structure, accessible surface
	area, and main-chain torsional angles by deep neural networks.
	\newblock In {\em Prediction of protein secondary structure}, pages 55--63.
	Springer, 2017.
	
	\bibitem{liu2017eses}
	Beibei Liu, Bao Wang, Rundong Zhao, Yiying Tong, and Guo-Wei Wei.
	\newblock Eses: software for e ulerian solvent excluded surface, 2017.
	
	\bibitem{dolinsky2004pdb2pqr}
	Todd~J Dolinsky, Jens~E Nielsen, J~Andrew McCammon, and Nathan~A Baker.
	\newblock Pdb2pqr: an automated pipeline for the setup of poisson--boltzmann
	electrostatics calculations.
	\newblock {\em Nucleic acids research}, 32(suppl\_2):W665--W667, 2004.
	
	\bibitem{case2008amber}
	David~A Case, Tom~A Darden, Thomas~E Cheatham, Carlos~L Simmerling, Junmei
	Wang, Robert~E Duke, Ray Luo, MRCW Crowley, Ross~C Walker, Wei Zhang, et~al.
	\newblock Amber 10.
	\newblock Technical report, University of California, 2008.
	
	\bibitem{brooks2009charmm}
	Bernard~R Brooks, Charles~L Brooks~III, Alexander~D Mackerell~Jr, Lennart
	Nilsson, Robert~J Petrella, Beno{\^\i}t Roux, Youngdo Won, Georgios
	Archontis, Christian Bartels, Stefan Boresch, et~al.
	\newblock Charmm: the biomolecular simulation program.
	\newblock {\em Journal of computational chemistry}, 30(10):1545--1614, 2009.
	
	\bibitem{chen2011mibpb}
	Duan Chen, Zhan Chen, Changjun Chen, Weihua Geng, and Guo-Wei Wei.
	\newblock Mibpb: a software package for electrostatic analysis.
	\newblock {\em Journal of computational chemistry}, 32(4):756--770, 2011.
	
	\bibitem{nguyen2019agl}
	Duc~Duy Nguyen and Guo-Wei Wei.
	\newblock {AGL-Score: A}lgebraic graph learning score for protein--ligand
	binding scoring, ranking, docking, and screening.
	\newblock {\em Journal of chemical information and modeling}, 59(7):3291--3304,
	2019.
	
	\bibitem{srivastava2014dropout}
	Nitish Srivastava, Geoffrey Hinton, Alex Krizhevsky, Ilya Sutskever, and Ruslan
	Salakhutdinov.
	\newblock Dropout: a simple way to prevent neural networks from overfitting.
	\newblock {\em The journal of machine learning research}, 15(1):1929--1958,
	2014.
	
	\bibitem{eua2}
	{FACT SHEET FOR HEALTH CARE PROVIDERS EMERGENCY USE AUTHORIZATION (EUA) OF
		REGEN-COV (fda.gov)}.
	
	\bibitem{weisblum2020escape}
	Yiska Weisblum, Fabian Schmidt, Fengwen Zhang, Justin DaSilva, Daniel Poston,
	Julio~CC Lorenzi, Frauke Muecksch, Magdalena Rutkowska, Hans-Heinrich
	Hoffmann, Eleftherios Michailidis, et~al.
	\newblock Escape from neutralizing antibodies by {SARS-CoV-2} spike protein
	variants.
	\newblock {\em Elife}, 9:e61312, 2020.
	
	\bibitem{wang2021antibody}
	Pengfei Wang, Manoj~S Nair, Lihong Liu, Sho Iketani, Yang Luo, Yicheng Guo,
	Maple Wang, Jian Yu, Baoshan Zhang, Peter~D Kwong, et~al.
	\newblock Antibody resistance of {SARS-CoV-2} variants {B}. 1.351 and {B}. 1.1.
	7.
	\newblock {\em Nature}, 10, 2021.
	
	\bibitem{planas2021reduced}
	Delphine Planas, David Veyer, Artem Baidaliuk, Isabelle Staropoli, Florence
	Guivel-Benhassine, Maaran~Michael Rajah, Cyril Planchais, Fran{\c{c}}oise
	Porrot, Nicolas Robillard, Julien Puech, et~al.
	\newblock Reduced sensitivity of {SARS-CoV-2} variant delta to antibody
	neutralization.
	\newblock {\em Nature}, pages 1--7, 2021.
	
\end{thebibliography}
% \end{multicols}
%\printbibliography

\newpage
\title{Supporting information for \\ Omicron BA.2 (B.1.1.529.2): high potential to becoming the next dominating variant}
\maketitle
\beginsupplement
\section{Supplementary figures}\label{Sfigures}
\begin{figure}[ht!]
	\centering
	\includegraphics[width = 0.8\textwidth]{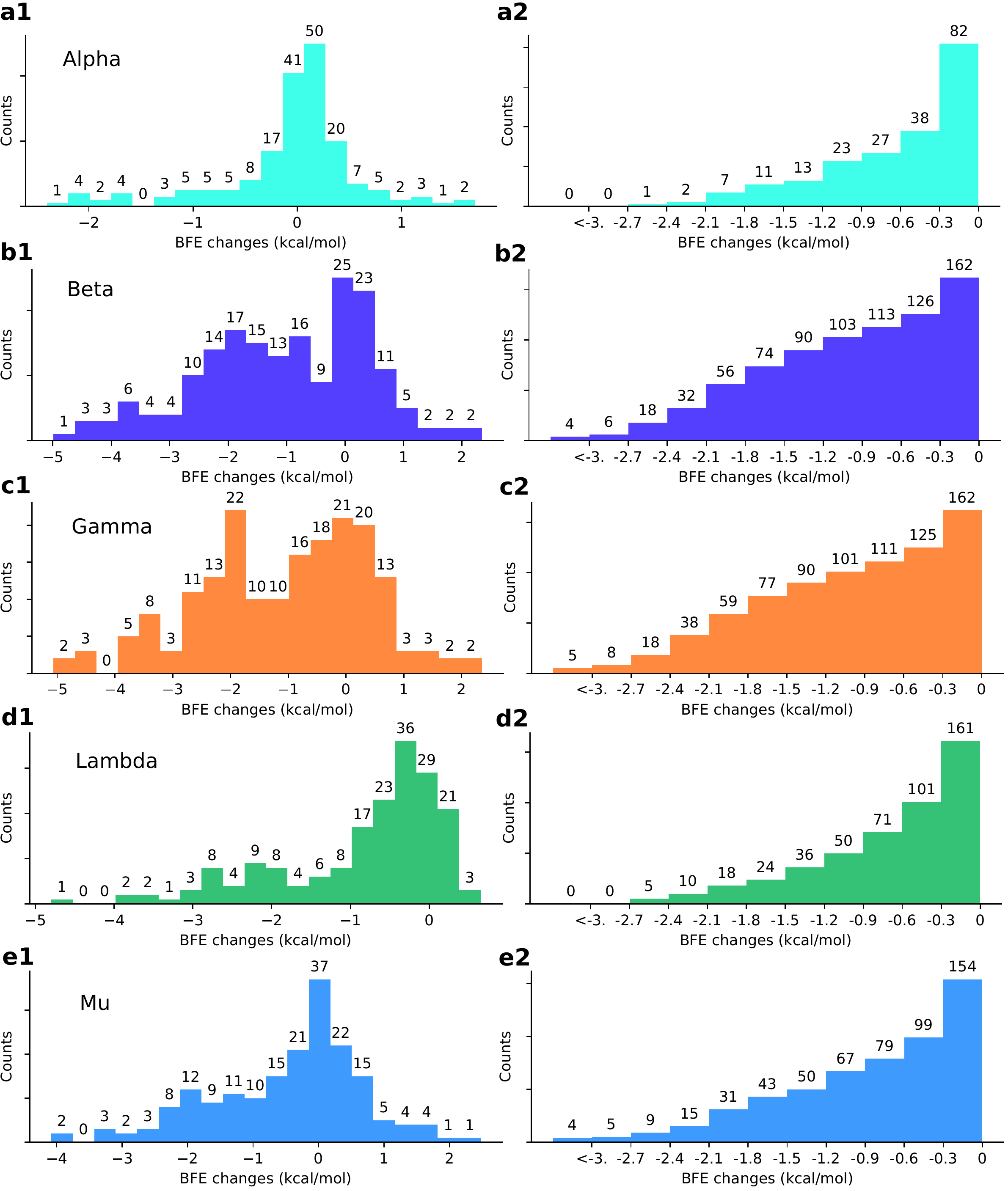}
	\caption{The extension of Figure~\ref{fig:combine2_2}. Analysis of variant mutation-induced BFE changes of 185 antibody-RBD complexes.   
		{\bf a1, b1, c1, d1} and {\bf e1}  The distributions (counts) of accumulated BFE changes  induced by Alpha, Beta, Gamma, Lambda, and Mu  mutations respectively for 185 antibody-RBD complexes. For each case, there are more mutation-weakened complexes than mutation-strengthened  complexes.
		{\bf a2, b2, c2, d2} and {\bf e2}  The numbers  of antibody-RBD complexes regarded as disrupted by  Alpha, Beta, Gamma, Lambda, and Mu   mutations respectively   under different thresholds ranging from 0 kcal/mol, -0.3 kcal/mol, to $<$-3 kcal/mol. }
	\label{fig:combine_SI_4}
\end{figure}
Figure~\ref{fig:combine_SI_4} provides the statistic analysis of BFE changes of RBD-ACE2 induced by mutations from Alpha, Beta, Gamma, Lambda, and Mu.

\section{Supplementary data}\label{Sdata}
The Supplementary\_Data.zip contains four files as listed in the following subsection.

\subsection{Disrupted antibodies}
File antibodies\_BFEs.csv shows the BFE changes of antibodies disrupted by Omicron mutations.

\subsection{List of antibodies}
File antibodies.csv lists the Protein Data Bank (PDB) IDs for all of the 185 SARS-CoV-2 antibodies.

\section{Supplementary feature generation methods}\label{Sfeature}
\begin{figure}[ht!]
	\setlength{\unitlength}{1cm}
	\begin{center}
		\includegraphics[width = 0.7\textwidth]{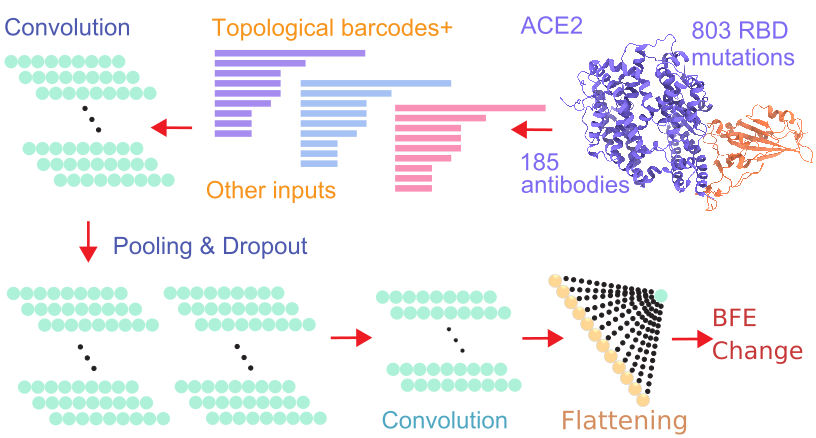}
		\caption{Illustration of genome sequence data pre-processing and BFE change predictions.}
		\label{fig_combine_all}
	\end{center}
\end{figure}
In this section, the workflow of the deep learning-based BFE change predictions of protein-protein interactions induced by mutations for the present SARS-CoV-2 variant analysis and prediction will be firstly introduced, which includes four steps as shown in \autoref{fig_combine_all}: (1) training data preparation; (2) feature generations of protein-protein interaction complexes; and (3) prediction of protein-protein interactions by deep neural networks. Next, the validation of our machine learning-based model will be demonstrated, suggesting consistent and reliable results compared to the experimental deep mutations data.

\subsection{Methods for BFE change predictions}
In this section, the process of the machine learning-based BFE change predictions is introduced. Once the data pre-processing and SNP genotyping are carried out, we will firstly proceed with the training data preparation process, which plays a key role in reliability and accuracy. A library of 185 antibodies and RBD complexes, as well as an ACE2-RBD complex, are obtained from Protein Data Bank (PDB). RBD mutation-induced BFE changes of these complexes are evaluated by the following machine learning model. According to the emergency and the rapid change of RNA virus, it is rare to have massive experimental BFE change data of SARS-CoV-2, while, on the other hand, next-generation sequencing data is relatively easy to collect. In the training process, the dataset of BFE changes induced by mutations of the SKEMPI 2.0 dataset \cite{jankauskaite2019skempi} is used as the basic training set, while next-generation sequencing datasets are added as assistant training sets. The SKEMPI 2.0 contains 7,085 single- and multi-point mutations and 4,169 elements of that in 319 different protein complexes used for the machine learning model training. The mutational scanning data consists of experimental data of the binding of ACE2 and RBD induced mutations on ACE2\cite{procko2020sequence} and RBD\cite{starr2020deep, linsky2020novo}, and the binding of CTC-445.2 and RBD with mutations on both protein\cite{linsky2020novo}.

Next, the feature generations of protein-protein interaction complexes are performed. The element-specific algebraic topological analysis on complex structures is implemented to generate topological bar codes \cite{chen2021revealing, carlsson2009topology,edelsbrunner2000topological,xia2014persistent}. In addition, biochemistry and biophysics features such as Coulomb interactions, surface areas, electrostatics, et al., are combined with topological features \cite{chen2021prediction}. The detailed information about the topology-based models will be demonstrated in \autoref{sec:topology}. Lastly, deep neural networks for SARS-CoV-2 are constructed for the BFE change prediction of protein-protein interactions \cite{chen2021revealing}. The detailed descriptions of dataset and machine learning model are found in the literature \cite{chen2020mutations,wang2020mutations,chen2021revealing} and are available at \href{https://github.com/WeilabMSU/TopNetmAb}{TopNetmAb}.

\subsection{Feature generation for machine learning model}\label{sec:topology}

\subsubsection{Topology features}
Among all features generated for machine learning prediction, the application of topology theory makes the model to a whole new level. Those summarized as other inputs are called as auxiliary features and are described in Section S3.3.2 and S3.3.3. In this section, a brief introduction about the theory of topology will be discussed. Algebraic topology \cite{carlsson2009topology,edelsbrunner2000topological} has achieved tremendous success in many fields including biochemical and biophysical properties\cite{xia2014persistent}. Special treatment should be implemented for biology applications to describe element types and amino acids in poly-peptide mathematically, which element-specific and site-specific persistent homology \cite{wang2020topology,chen2020mutations}. To construct the algebraic topological features on protein-protein interaction model, a series of element subsets for complex structures should be defined, which considers atoms from the mutation sites, atoms in the neighborhood of the mutation site within a certain distance, atoms from antibody binding site, atoms from antigen binding site, and atoms in the system that belong to type of \{C, N, O\}, $\mathcal{A}_\text{ele}(\text{E})$. Under the element/site-specific construction, simplicial complexes is constructed on point clouds formed by atoms. For example, a set of independent $k\!+\!1$ points is from one element/site-specific set $U=\{u_0, u_1, ...,u_k\}$. The $k$-simplex $\sigma$ is a convex hull of $k\!+\!1$ independent points $U$, which is a convex combination of independent points. For example, a $0$-simplex is a point and a $1$-simplex is an edge. Thus, a $m$-face of the $k$-simplex with $m\!+\!1$ vertices forms a convex hull in a lower dimension $m<k$ and is a subset of the $k\!+\!1$ vertices of a $k$-simplex, so that  a sum of all its $(k\!-\!1)$--faces is the boundary of a $k$--simplex $\sigma$ as
\begin{equation}
\partial_k\sigma = \sum_{i=1}^{k}(-1)^i\langle u_0, ..., \hat{u}_i, ..., u_k\rangle ,
\label{eq_boundary_operator}
\end{equation}
where $\langle u_0, ..., \hat{u}_i, ..., u_k\rangle $ consists of all vertices of $\sigma$ excluding $u_i$. The collection of finitely many simplices is a simplicial complex. In the model, the Vietoris-Rips (VR) complex (if and only if $\mathbb{B}(u_{i_j}, r)\cap\mathbb{B}(u_{i_{j'}}, r)\ne\emptyset$ for ${j}, {j'} \in [0,k]$) is for dimension 0 topology, and alpha complex (if and only if $\cap_{u_{i_j}\in\sigma}\mathbb{B}(u_{i_j}, r)\ne\emptyset$) is for point cloud of dimensions 1 and 2 topology \cite{xia2014persistent}. 

The $k$-chain $c_k$ of a simplicial complex $K$ is a formal sum of the $k$-simplices in $K$, which is $c_k=\sum\alpha_i\sigma_i$, where  $\alpha_i$ is coefficients and is chosen to be $\mathbb{Z}_2$. Thus, the boundary operator on a $k$-chain $c_k$ is
\begin{equation}
\partial_kc_k=\sum\alpha_i\partial_k\sigma_i,
\label{eq_boundary_operator_chain}
\end{equation}
such that $\partial_k:C_k\rightarrow C_{k-1}$ and follows from that boundaries are boundaryless $\partial_{k-1}\partial_k=\emptyset$. 
A chain complex is
\begin{equation}
\cdots \stackrel{\partial_{i+1}}\longrightarrow C_i(K) \stackrel{\partial_i}\longrightarrow C_{i-1}(K) \stackrel{\partial_{i-1}}\longrightarrow \cdots \stackrel{\partial_2} \longrightarrow C_{1}(K) \stackrel{\partial_{1}}\longrightarrow C_0(K) \stackrel{\partial_0} \longrightarrow 0,
\label{eq_chain_complex}
\end{equation}
as a sequence of complexes by boundary maps. Therefore, the Betti numbers are given as the ranks of $k$th homology group $H_k$ as $\beta_k=\text{rank}(H_k)$, where $H_k=Z_k/B_k$, $k$-cycle group $Z_k$ and the $k$-boundary group $B_k$. The Betti numbers are the key for topological features, where $\beta_0$ gives the number of connected components, such as number of atoms, $\beta_1$ is the number of cycles in the complex structure, and $\beta_2$ illustrates the number of cavities. This presents abstract properties of the 3D structure.

Finally, only one simplicial complex couldn't give the whole picture of the protein-protein interaction structure. A filtration of a topology space is needed to extract more properties.
A filtration is a nested sequence such that
\begin{equation}
\emptyset = K_0 \subseteq K_1 \subseteq \cdots \subseteq K_m = K.
\label{eq_filtration}
\end{equation}
Each element of the sequence could generate the Betti numbers $\{\beta_0, \beta_1, \beta_2\}$ and consequentially, a series of Betti numbers in three dimensions is constructed and applied to be the topological fingerprints in Figure~\ref{fig_combine_all}.

% The feature generation of protein-protein interactions is crucial for BFE change predictions induced by mutations. In the main content, we briefly discuss the topology theory and its implementation in this model. There are other features, named as auxiliary features, which also play important roles in the training process. The auxiliary features includes chemical and physical information of the complexes, such as molecular surface areas, partial charges, Coulomb interactions, van der Waals interaction, mutation site neighborhood amino acid composition, pKa shifts, electrostatic solvation free energy, and secondary structure information\cite{wang2020topology,chen2021prediction}. Two different levels of features which are residue-level and atom-level and will be discussed following.

\subsubsection{Residue-level features}

{\bf Mutation site neighborhood amino acid composition} 
Neighbor residues are the residues within 10 {\AA } of the mutation site. Distances between residues are calculated based on residue C$_{\alpha}$ atoms. Six categories of amino acid residues are counted, which are hydrophobic, polar, positively charged, negatively charged, special cases, and pharmacophore changes. The count and percentage of the 6 amino acid groups in the neighbor site are regrading as the environment composition features of the mutation site. The sum, average, and variance of residue volumes, surface areas, weights, and hydropathy scores are used but only the sum of charges is included. 

{\bf pKa shifts} 
The pKa values are calculated by the PROPKA software \cite{bas2008very}, namely the values of 7 ionizable amino acids, namely, ASP, GLU, ARG, LYS, HIS, CYS, and TYR. The maximum, minimum, sum, the sum of absolute values, and the minimum of the absolute value of total pKa shifts are calculated. 
We also consider the difference of pKa values between a wild type and its mutant. 
Additionally, the sum and the sum of the absolute value of pKa shifts based on ionizable amino acid groups are included.

{\bf Position-specific scoring matrix (PSSM)}
Features are computed from the conservation scores in the position-specific scoring matrix of the mutation site  for the wild type and the mutant as well as their difference. The conservation scores are generated by PSI-BLAST \cite{altschul1997gapped}.

{\bf Secondary structure} 
The SPIDER2 software is used to compute the probability scores for residue  
torsion angle and residues being in a coil, alpha helix, and  beta strand
based on the sequences for the wild type and the mutant \cite{yang2017spider2}.

\subsubsection{Atom-level features}
Seven groups of atom types, including C, N, O, S, H, all heavy atoms, and all atoms, are considered when generating the element-type features. Meanwhile, other three atom types, i.e., mutation site atoms, all heavy atoms, and all atoms, are used when generating the general atom-level features.

{\bf Surface areas} Atom-level solvent excluded surface areas are computed by ESES \cite{liu2017eses}.

{\bf Partial changes} Partial change of each atom is generated by pdb2pqr software \cite{dolinsky2004pdb2pqr} using the  Amber force field \cite{case2008amber} for wild type and CHARMM force field \cite{brooks2009charmm} for mutant. The sum of the partial charges and the sum of absolute values of partial charges for each atomic group are collected.

{\bf Atomic pairwise interaction interactions} 
Coulomb energy of the $i$th single atom is calculated as the sum of pairwise coulomb energy with every other atom as
\begin{equation}
C_i=\sum_{j,j\ne i}k_e \frac{q_iq_j}{r_{ij}},
\end{equation}
where $k_e$ is the Coulomb's constant, $r_{ij}$ is the distance of $i$th atom to $j$th atom, and $q_i$ is the charge of $i$th atom. The van der Waals energy of the $i$th atom is modeled as the sum of pairwise Lennard-Jones potentials with other atoms as
\begin{equation}
V_i=\sum_{j,j\ne i}\epsilon\Big[\big(\frac{r_i+r_j}{r_{ij}}\big)^{12}-2\big(\frac{r_i+r_j}{r_{ij}}\big)^{6}\Big],
\end{equation}
where $\epsilon$ is the depth of the potential well, and $r_i$ is van der Waals radii.

In atomic pairwise interaction, 5 groups (C, N, O, S, and all heavy atoms) are counted both for Coulomb interaction energy and van der Waals interaction energy.

{\bf Electrostatic solvation free energy} 
Electrostatic solvation free energy of each atom is calculated using the Poisson-Boltzmann equation via MIBPB \cite{chen2011mibpb} and are summed up by atom groups.

\section{Supplementary machine learning methods}\label{Smethods}
The topology-based network model for BFE change predictions induced mutations on SARS-CoV-2 studying applies a deep neural network structure. Similar approaches have been widely implemented in the energy prediction of protein-ligand binding\cite{nguyen2019agl} and protein-protein interactions\cite{wang2020topology}. The neural network method maps an input feature layer to output layer and mimics biological brains for solving problems where numerous neuron units are involved and weights of neurons are updated by backpropagation methods. To make more complicated structure in order to extract abstract properties, more layers and more neurons in each layer can be constructed. In the training process, optimization methods are applied to avoid overfitting issue, such as dropout methods\cite{srivastava2014dropout} that a partial of computed neurons of each layer is dropped. For the model cross validations, the Pearson correlation of 10-fold cross-validation is 0.864, and the root mean square error is 1.019 kcal/mol. 

\subsection{Deep learning algorithms}
A deep neural network is a neural network method with multi-layers (hidden layers) of neurons between the input and output layers. In each layer, the single neuron gets fully connected with the neurons in the next layer. It should preserve the consistency of all labels when applying the model for mutation-induced BFE change predictions. The loss function is constructed as follows:

\begin{eqnarray}
\mathop{\text{argmin}}_{{W}, {b}}L({W}, {b}) = \mathop{\text{argmin}}_{{W}, {b}} \frac{1}{2}\sum_{i=1}^{N}\big(y_i-f({x}_i; \{{W}, {b}\})\big)^2+\lambda\|{W}\|^2
\label{eq_loss_fcn}
\end{eqnarray}
where $N$ is the number of samples, $f$ is a function of the feature vector $x_i$ parameterized by a weight vector $W$ and bias term $b$, and $\lambda$ represents a penalty constant.

\subsection{Optimization}
The backpropagation is applied to evaluate the loss function starting from the output layer and propagates backward through the network structure to update the weight vector $W$ and bias term $b$. Since gradient calculation is required, therefore, we apply the stochastic gradient descent method with momentum, which only evaluates a small part of training data and can be considered as calculating exponentially weighted averages, which is given as 
\begin{equation}
\begin{split}
&V_i = \beta V_{i-1} + \eta \nabla_{W_i}L(W_i, b_i) \\
&W_{i+1} = W_i - V_i,
\end{split}
\end{equation}
where $W_i$ is the parameters in the network, $L(W_i,b_i)$ is the objective function, $\eta$ is the learning rate, ${X}$ and ${y}$ are the input and target of the training set, and $\beta\in [0,1]$ is a scalar coefficient for the momentum term.
The momentum term involved accelerates the converging speed. 

\section{Supplementary validation}\label{Svalidation}
In the main content, we briefly summarized validations of our machine learning predictions and experimental data. For large quantitative validations, we compared the BFE change prediction for mutations on S protein RBD to the experimental deep mutational enrichment data on RBD binding to human ACE2 and CTC-445.2 induced by RBD mutations \cite{chen2021revealing, chen2021prediction, linsky2020novo}. To make these validations, we eliminated the experimental deep mutational enrichment data of RBD binding to human ACE2 and CTC-445.2 from the training sets and set them as testing sets, which have 1539 and 1500 samples, respectively. In the validation of RBD and CTC-445.2 complex, there is a very high correlation between the enrichment data and machine learning predictions, as well as the validation of RBD binding to ACE2, with Pearson correlations are 0.69 and 0.70, respectively. The deep mutational enrichment data can give a proportional descriptor of the affinity strength of protein-protein interactions induced by mutations. The machine learning methods, however, give stable and equalized evaluations, while experimental data might be different dramatically due to conditions and environments.

In addition, we compared our machine learning results with other experimental data, which are escape fraction, pseudovirus infection changes, and IC$_{50}$ fold changes \cite{chen2021revealing}. In the comparison of 35 cases to experimental escape fractions on RBD binding to clinical trial antibodies induced by emerging mutations, our machine learning predictions have a Pearson correlation of 0.80. Especially, those high escaping mutations E484K and E484Q on LY-CoV555, and mutations K417T and K417N on LY-CoV016, are indicated by both our predictions and the experimental data \cite{chen2021revealing}. We also use the pattern comparisons of our prediction to experimental data. Lastly, we collected experimental data from different literature \cite{eua2,weisblum2020escape,wang2021antibody, planas2021reduced}. According to variations from different research groups, they were summarized in increasing/decreasing patterns of emerging variant (including co-mutations) impacts on antibody therapies in clinical trials. In total, there are 20 pattern comparisons with an excellent agreement between various experimental data and our predictions, except for a minor discrepancy \cite{chen2021revealing}.

\end{document}